\documentstyle[12pt,aaspp4]{article}

\slugcomment{to appear in ApJ Letters}  

\begin{document}

\title{VLA Imaging of the Disk Surrounding the Nearby Young Star TW Hya}

\author{D.J. Wilner and  P.T.P. Ho}
\affil{Harvard-Smithsonian Center for Astrophysics, 60 Garden Street,
    Cambridge, MA 02138}

\author{J.H. Kastner}
\affil{Rochester Institute of Technology, Chester F. Carlson Center for 
Imaging Science, 54 Lomb Memorial Drive, Rochester, NY 14623}

\and
\author{L.F. Rodr\'\i guez}
\affil{Instituto de Astronom\'\i a, UNAM, Apdo. Postal 70-264,
04510 M\'exico, D.F., M\'exico}

\begin{abstract}
The TW Hya system is perhaps the closest analog to the early solar nebula.
We have used the Very Large Array to image TW Hya at wavelengths 
of 7~mm and 3.6~cm with resolutions $0\farcs1$ ($\sim5$ AU) 
and $1\farcs0$ ($\sim50$ AU), respectively.
The 7~mm emission is extended and appears dominated by a dusty disk of 
radius $>50$~AU surrounding the star. The 3.6~cm emission is unresolved and 
likely arises from an ionized wind or gyrosynchrotron activity.
The dust spectrum and spatially resolved 7~mm images of the TW Hya disk are 
fitted by a simple model with temperature and surface density described by 
radial power laws, $T(r)\propto r^{-0.5}$ and $\Sigma(r) \propto r^{-1}$.
These properties are consistent with an irradiated gaseous 
accretion disk of mass $\sim0.03~{\rm M_{\odot}}$ with an accretion rate 
$\sim10^{-8}~{\rm M_{\odot}~yr^{-1}}$ and viscosity parameter 
$\alpha = 0.01$. The estimates of mass and mass accretion rates are 
uncertain as the gas-to-dust ratio in the TW Hya disk may have evolved 
from the standard interstellar value.

\end{abstract}

\keywords{stars: individual (TW Hydrae) --- stars: formation ---
circumstellar matter --- radio continuum: stars --- accretion: accretion disks}

\section{Introduction}
The TW Hya association, which consists of at least 13 pre-main-sequence 
systems at a distance of about 50~pc, is thought to be the nearest region 
of recent star formation (Kastner et al. 1997, Webb et al. 1999a,b). 
The age of the association
members has been estimated by various techniques at 5 to 20 million years,
an important time period for the formation of planetary systems 
(Kastner et al. 1997, Jura et al. 1998, Soderblom et al. 1998,
Webb et al. 1999b, Weintraub et al. 2000).
Because of its close proximity and interesting age, the association has 
proved to be fertile ground for observations of phenomena related to 
the structure and evolution of circumstellar disks.
Four of the association members were detected by {\em IRAS} indicating 
the presence of circumstellar dust, either precursors to the assemblage 
of larger bodies or perhaps the resulting debris. Recent observations with
better angular resolution toward members of this association
have revealed a disk around TW Hya visible in scattered near-infrared light 
(Weinberger et al. 1999) and detectable in 
various molecular tracers (Zuckerman et al. 1995, Kastner et al. 1997),
a dust disk around HR 4796A with a cleared interior gap (Koerner et al. 1998, 
Jayawardhana et al. 1998, Schneider et al. 1999), a dust disk within the 
hierarchical quadruple system HD~98800 (Low et al. 1999), and direct images 
of a substellar companion to CoD~-33$^{\circ}$~7795 (Lowrance et al. 1999,
Webb et al. 1999b). 

Among the TW Hya association members, TW Hya, a K7 star 
of $\sim0.5~M_{\odot}$, has the strongest circumstellar dust 
emission and a disk mass perhaps comparable to the solar nebula prior 
to the formation of the planets. The TW Hya system is almost three times 
closer than the young stars in nearby dark clouds like Taurus, Ophiucus and 
Chameleon, with a distance of $56\pm7$~pc determined by {\em Hipparcos}. 
The technique of millimeter interferometry can now probe emission from the 
dusty disk material at subarcsecond scales, comparable to the best resolution 
achieved at optical and infrared wavelengths (Mundy et al. 1996, 
Wilner et al. 1996, Dutrey et al. 1996, Wilner \& Lay 2000).  
Dust emission at long millimeter wavelengths is nearly entirely optically thin 
and arises from the full disk column, including the innermost regions of 
planet formation. 
In this {\em Letter}, we describe observations of thermal emission from 
TW Hya obtained with the Very Large Array (VLA)
of the National Radio Astronomy Observatory
\footnote{The National Radio Astronomy Observatory is a facility of the 
National Science Foundation operated under cooperative agreement by 
Associated Universities, Inc.}.
These observations directly resolve the dusty disk surrounding the star
and provide constraints on its structure. 

\section{Observations}
We observed TW Hya with the VLA on two occasions in excellent weather
conditions, first in a short observation in February 1998 in the 
D configuration as part of larger survey of dust in T Tauri systems, and 
then for a full track in October 1999 in the BnA configuration to obtain 
higher angular resolution. 
The VLA was divided into two subarrays in each observing session, 
one consisting of those antennas equipped with 7~mm receivers, and 
the other consisting of the remaining antennas, which were set to observe
at 3.6~cm. Table~\ref{tab:vlaobs} summarizes the observational parameters. 
All observations were made with the maximum bandwidth, two 50~MHz channels 
in each of two circular polarizations, for the best continuum sensitivity.
For the observations in the BnA configuration, phase calibration at 7~mm 
was accomplished by rapid phase referencing within a 120 second cycle
to the nearby calibrator 1037-295.
The absolute flux scale was determined through observations of the standard 
source 3C286 and should be accurate to 10\%. All data calibration and imaging 
were performed using the standard routines in the AIPS software package. 

\section{Results}

\subsection{7~mm}
\begin{figure}
\plotfiddle{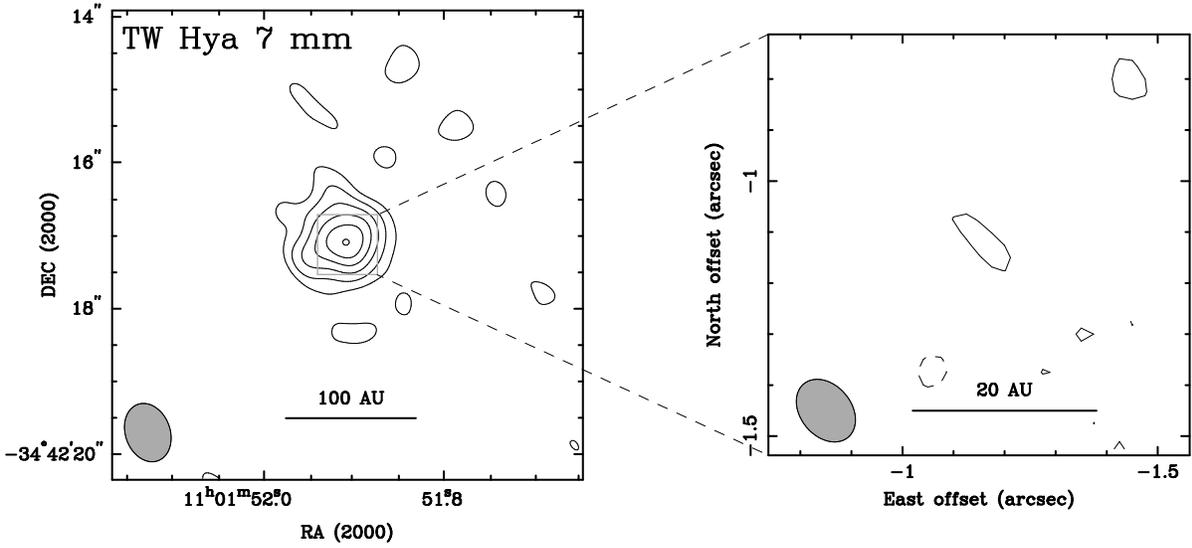}{2.5in}{270}{70}{70}{-270}{210}
\caption{VLA 7~mm images of TW Hya at two resolutions.
The synthesized beams are $0\farcs82\times0\farcs61$ p.a. 20 (left panel)
and $0\farcs13\times0\farcs10$, p.a. 39 (right panel). Contour levels
are $\pm(2,3,4,5,6)\times$ the noise levels of 0.5 and 0.4 mJy/beam,
which correspond to brightness temperatures 0.66 and 20 K.
Negative contours are dashed.
}
\label{fig:Qband}
\end{figure}

Figure~\ref{fig:Qband} shows images made from the 7~mm data at two different 
resolutions obtained with different visibility weighting schemes.
The lower resolution image in the left panel of Figure~\ref{fig:Qband}
was made with natural weighting
and a 300~k$\lambda$ Gaussian taper to produce a $\sim0\farcs6$ beam that
emphasizes the spatially extended low brightness emission. The region of 
detectable 7~mm emission at this resolution is $\sim100$~AU in diameter. 
The total flux 
integrated within a box of size $2''$ is $8\pm1$ mJy. It is possible
that a fainter halo extends to even larger distances from the peak.
The emission presents a roughly circular boundary, consistent with other
suggestions that the emission arises in a disk oriented close to pole-on
(Kastner et al. 1997), though the ellipticity is uncertain and does not 
provide a strong constraint on the inclination.
The image in the right panel of Figure~\ref{fig:Qband} was made with robust 
weighting to obtain a $\sim0\farcs1$ beam (5.6 AU), close to the highest 
resolution available from the longest baselines in these data.  Little 
detectable emission is visible from TW~Hya at this scale, though a weak signal
remains at the center of the larger structure, with perhaps a comparable peak
offset to the northwest.  Note that while the flux sensitivity of the higher 
resolution image is a little better than that of the lower resolution image, 
the brightness temperature sensitivity is considerably worse because of 
the smaller synthesized beam. 

\subsection{3.6~cm}
TW Hya was detected at 3.6~cm as an unresolved source with 
flux density $0.20\pm0.028$ mJy. 
(The results from the two epochs are consistent within the noise.)
The emission peak is located at $11^{\rm h}01^{\rm m}51\fs91$, 
$-34^{\circ}42'16\farcs96$ (J2000) with an estimated uncertainty of 
$\pm0\farcs1$. 
Rucinski (1992) previously used the VLA to search for radio emission
from TW Hya at 3.6, 6 and 20~cm, without success. He reports an rms noise 
level of 0.028 mJy at 3.6~cm, very similar to the value obtained here 
with fewer antennas and more integration time, and so TW Hya
should have been detectable in the earlier observation (with a synthesized
beam 15 times larger in area).  If the observation of Rucinski (1992) 
is correct, then the radio emission from TW~Hya must be time variable. 

\section{Discussion}

\subsection{Disk Spectrum}
\begin{figure}
\plotfiddle{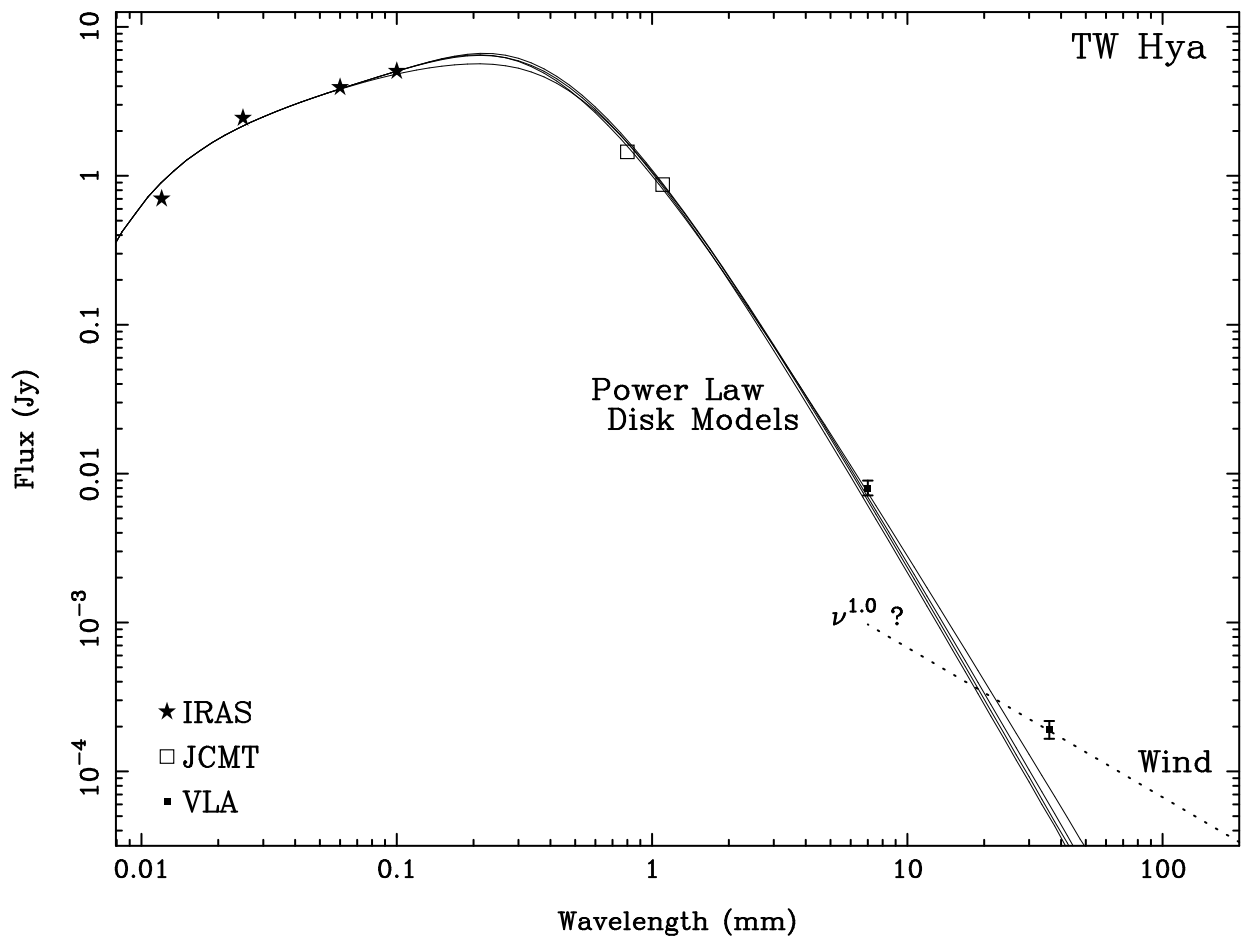}{4in}{270}{70}{70}{-260}{410}
\caption{Long wavelength spectrum of TW Hya, including observations 
in the infrared from {\em IRAS}, 
the submillimeter from the JCMT (Weintraub et al. 1989) 
and the radio from the VLA (this paper).
Solid lines show best fit power law disk models with values of the 
surface density power law index of 0, 0.5, 1 and 1.5
(left to right solid lines in the bottom right part of the figure).
The dotted line indicates a possible ionized wind component 
with a large positive spectral index (see text).
}
\label{fig:sed}
\end{figure}

Figure~\ref{fig:sed} shows the broadband spectrum of TW~Hya from 
mid-infrared to radio wavelengths. Like many T Tauri stars, the long
wavelength emission, far in excess of the stellar photosphere, 
is well fitted by a family of thin disk models parameterized by radial 
power laws in temperature and surface density 
(Adams, Lada \& Shu 1988, Beckwith et al. 1990). In these models, 
the slope of the spectrum in the infrared, where the disk is optically thick, 
constrains the temperature distribution. At millimeter wavelengths, where 
the disk is largely optically thin, the emission is proportional to the disk 
mass weighted by the temperature distribution. Irradiation from the star 
and low optical depths for the outer disk, together with flaring, tend 
to drive $T(r)\sim r^{-0.5}$ (Kenyon \& Hartmann 1987, D'Alessio et al. 1998).
For illustration, Figure~\ref{fig:sed} shows spectra from a series of 
face-on disk models with outer radius 100~AU and the usual (constant) 
dust opacity law 
$\kappa = 0.1 (\lambda/250~\mu m)^{-\beta}$~cm$^{2}$~g$^{-1}$, $\beta=1$, 
and $\Sigma(r) \propto (r/1~{\rm AU})^{-p}$~g~cm$^{-2}$,
$p=0, 0.5, 1.0$, and $1.5$ (with the mass of gas+dust adjusted from 
0.044 to 0.034 ${\rm M_{\odot}}$ to provide the best least squares fit)
for $T(r) = 110 (r/1~{\rm AU})^{-q}$~K, $q=0.55$.
Note that the spectrum is not very sensitive to $p$, the power law index 
of the surface density distribution.

Uncertainties in mass opacity coefficient dominate the uncertainties
in the estimate of disk mass as the standard value derived from interstellar
clouds may not apply (see the review by Beckwith et al. 2000). In addition to 
issues of grain composition, size and shape, the standard coefficient assumes 
a gas-to-dust ratio of 100 by mass.
For TW Hya, Kastner et al. (1997) observe $^{13}$CO emission and calculate a 
gas mass of $3.5\times10^{-5}$~$M_{\odot}$, three orders of magnitude lower 
than implied by the millimeter continuum emission. One explanation for this 
commonly found discrepancy between tracers of total (gas+dust) disk mass 
is the spectre of severe molecular depletion (Dutrey et al. 1996). 
Another possibility is a dramatic decrease in the gas-to-dust ratio, which 
Zuckerman et al. (1995) argue may be especially appropriate for older systems 
like TW Hya where substantial processing of disk material may have occurred. 
The disk gas is lost on a timescale of $\sim10^{7}$~yr, perhaps incorporated
into giant planets, while dust apparently persists for $\sim10^{8}$~yr or more. 
For TW Hya, the strong accretion signatures at ultraviolet and optical 
wavelengths indicate the presence of some gas close to the star. 
But all estimates of the total disk mass, which is likely 
dominated by molecular hydrogen, remain somewhat problematic.

At 3.6~cm, emission from the disk models falls far short of the 
observed flux. By analogy with other young stellar objects, the 
3.6~cm emission may arise from hot plasma that originates in 
a stellar wind. The spectrum of this ionized component may be flat 
(if optically thin) or rising (if partially optically thick) 
with a spectral index perhaps as large as unity (Anglada et al. 1998). 
Alternatively, the 3.6~cm emission may be attributed to pre-main-sequence
magnetic activity, an especially attractive source if the emission
is time variable (Feigelson \& Montmerle 1999). 
The same activity could also account for the 
variable X-ray emission from the star (Kastner et al. 1999).
A dotted line in Fig~\ref{fig:sed} shows an extrapolation of the spectrum 
for a wind with spectral index unity, which maximally contributes to the
emission at shorter wavelengths; even in this case, the hot plasma 
contribution at 7~mm is 1~mJy, less than 15\% of the dust emission 
from the disk models.

\subsection{Disk Structure}
\begin{figure}
\plotfiddle{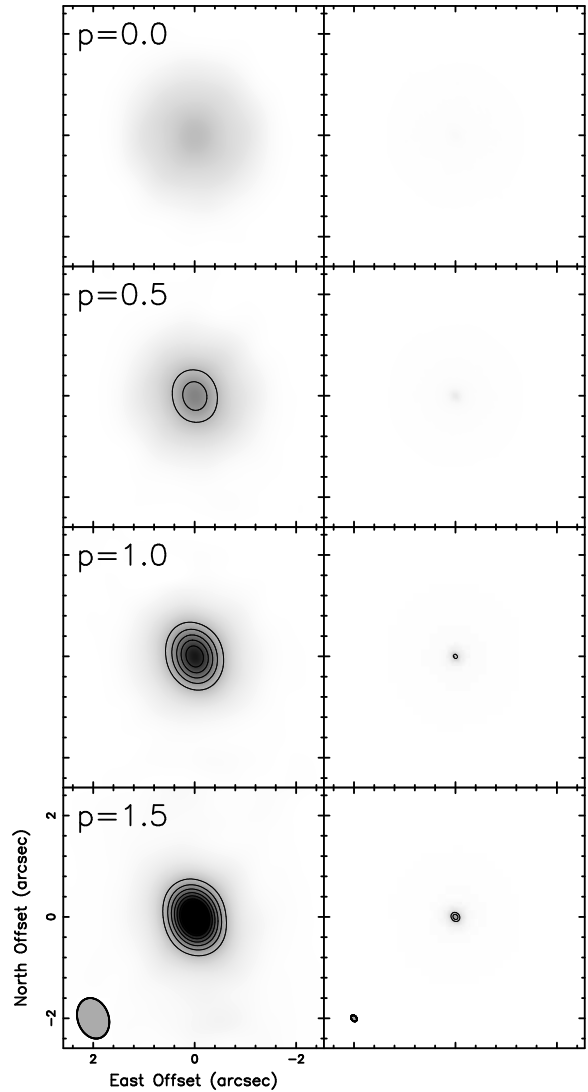}{5.5in}{0}{65}{65}{-200}{-30}
\caption{Simulated VLA images for the four model disks whose spectra 
are shown in Figure~\ref{fig:sed} made with the same angular resolutions
as the two images of TW Hya shown in Figure~\ref{fig:Qband}.
The model disks are face-on, sampled 
on the $u,v$ tracks obtained for TW Hya, and imaged in the standard way.
The contour levels match those in Figure~\ref{fig:Qband}. 
A logarithmic grey scale shows low brightness emission.
The left panels indicate the values of $p$, the power law exponent 
of the surface density distribution, which largely sets the central 
concentration of the disk emission.  The synthesized beams are 
illustrated in the lower left corners of the bottom panels.
}
\label{fig:r100}
\end{figure}

The resolved 7~mm images are sensitive to the degree of central concentration 
of the disk emission. The brightness in each beam is given by the product of
temperature and opacity, where the latter quantity is proportional to the 
surface density when the optical depth is low. With the available sensitivity,
the disk emission can be detected only where temperatures and optical depths 
are sufficiently high. 
The images offer hints of inhomogeneities in the 
disk emission, in particular along a southeast-northwest axis, but these 
features have marginal significance (less than twice the rms noise). 
We consider only axisymmetric models for the disk in deriving structural 
parameters.  Figure~\ref{fig:r100} shows images of a series of four disk 
models that follow the standard power law description and match the TW Hya 
spectrum but have different surface density distributions. 
To mimic observations, the models have been imaged from the $(u,v)$ tracks
obtained for TW Hya for two resolutions and deconvolved with the standard 
algorithms.  The models in Figure~\ref{fig:r100} show that disks with 
steep surface density distributions produce detectable emission at 
small radii at high resolution while flatter distributions do not. 
The disks in these models are oriented face on and have outer radii of 
100 AU, but disks with modest inclinations and outer radii from a 
few 10's to a few 100's of AU exhibit qualitatively similar behavior.

Comparison of the model images in Figure~\ref{fig:r100} with the images 
in Figure~\ref{fig:Qband} suggest that values of $p$ as large as $1.5$ 
or as low as $0.5$ are not compatible with the 7~mm data. A model with 
$p$ as low as $0.5$ can more closely match the observations if the disk is 
made sufficiently small, with outer radius $<50$~AU, which raises the 
surface density and the resulting brightness. But the radial extent of 
scattered light in the Hubble Space Telescope {\em NICMOS} and {\em WFPC2}
images of TW~Hya show that disk material extends well beyond 
this radius (Weinberger et al. 1999, Krist et al. 2000).  
The steeper power laws are especially in conflict with the observations 
if any of the emission visible at the highest resolution can be attributed 
to an ionized component. The preferred value of $p$ is near unity, 
where the power law model matches the 7~mm observations imaged 
at low and high resolution.
This result requires that the dust properties within the disk are not a 
strong function of radial distance from the star.

The TW Hya spectrum and resolved 7~mm images are generally consistent 
with an irradiated gaseous accretion disk 
(D'Alessio et al. 1997, 1998, Chiang \& Goldreich 1997)
that follows the Shakura--Sunyaev $\alpha$ prescription. 
For steady accretion, the surface density is given by 
$\Sigma = \dot{M}/{3\pi\nu}$ away from the boundaries, 
where $\nu = \alpha c_s H$ is the kinematic viscosity parameterized by a 
local velocity (the sound speed, $c_s$) and scale length 
(the scale height, $H$) and a dimensionless parameter ($\alpha$). 
At radii where the disk is optically thin to its own radiation, a condition 
that likely holds beyond a few AU for the TW Hya system, the disk becomes 
nearly vertically isothermal.  For a midplane temperature distribution 
characterized by 
$T_m\propto r^{-1/2}$ due to irradiation, the surface density distribution
can be approximated by $\Sigma \propto (r^{3/2}T_m)^{-1} \sim r^{-1}$, in
good agreement with the data. While inhomogeneities are likely present in 
the disk, and changes in disk composition and optical depth will modify 
the energy balance and structure close to the star, the overall structure 
of the TW Hya disk appears amenable to this simple power law description. 
Substituting the numerical factors in the expression for surface density gives
$ \Sigma(r) \approx 240~(\dot{M}/10^{-8}~{\rm M_{\odot}~yr^{-1}})
                     (\alpha/0.01)^{-1}(r/{\rm AU})^{-1}$ g~cm$^{-2}$, 
where the scalings for $\alpha$ and $\dot{M}$ are typical values
for pre-main-sequence accretion disks (Hartmann et al. 1998).
For the TW Hya disk, the surface density at 1~AU radius in the $p=1$ model 
shown in Figure~\ref{fig:r100} is 450~g~cm$^{-2}$, about a factor of two 
larger than the canonical value. If the accretion model is valid, then 
the agreement between the determinations of the surface density-- 
within a factor of a few-- provides support for the accuracy of the 
adopted mass opacity law at millimeter wavelengths. More stringent 
tests will be provided by independent measures of the accretion rate 
and resolved observations of the dust emission at additional wavelengths. 

One consequence of the modest brightness observed within 5~AU of TW Hya 
is that the surface density is unlikely be very much higher in this 
region of the disk than expected from an $r^{-1}$ extrapolation inward.  
A region of low viscosity and concomitant high surface density in the inner 
disk may be required by some mechanisms for accretion, planet formation, and 
also planet migration. For example, the layered accretion of Gammie (1996) 
piles up accreting mass in a ``dead zone'' where the magnetohydrodynamic 
instability does not operate very effectively. For TW Hya, there is no 
evidence for any such substantial mass reservoir at small radii, at least 
not at the present epoch. 

\section{Conclusions}
The TW Hya system provides perhaps the closest analog to the early solar nebula,
and VLA observations with subarcsecond resolution characterize conditions 
in the circumstellar environment at the size scales of giant planet formation.
The images of TW Hya at $\lambda=7$~mm show extended emission that 
we attribute to dust in a disk viewed close to pole-on.
The dust emission is best fit with a power law disk model having surface 
density falling off as $r^{-1}$, as shallower or steeper power laws produce 
too little or too much signal at high resolution, especially at radii $<10$~AU.
The observed structure conforms to that derived in self-consistent calculations 
of gaseous accretion disks with heating dominated by irradiation from the 
central star. The observations provide hints of substructure within the disk, 
but data with much better sensitivity will be required to confirm 
the reality of these features. This nearby disk system will be a prime target 
for the next generation of millimeter and submillimeter arrays.

\acknowledgments
This research was partially supported by NASA Origins of Solar Systems
Program grant NAG5-8195. LFR acknowledges the support of CONACyT, Mexico
We thank Lee Hartmann and Nuria Calvet for 
valuable discussions about irradiated accretion disks.

\clearpage

\begin{deluxetable}{lcc}
\tablewidth{400pt}
\tablecaption{VLA Observations of TW Hya}
\tablehead{ 
\colhead{} & \colhead{7~mm} & \colhead{3.6~cm} }
\startdata
Pointing center (J2000):  & \multicolumn{2}{c}{$\alpha=11^{h}01^{m}52\fs0$,
                               $\delta=-34^{\circ}42\farcm16\farcs0$} \\ 
Date (configuration):   & \multicolumn{2}{c}{1999 Oct 24 (BnA)} \\
~~~~No. antennas:       & 16 & 11 \\
~                       & \multicolumn{2}{c}{1998 Feb 03 (D)} \\
~~~~No. antennas:       & 12 & 15 \\
Primary flux calibrator: & 3C286 & 3C286 \\
~~~~Adopted flux:       & 1.5 Jy & 5.2 Jy\\
Phase calibrator:        & 1037-295 & 1037-295 \\
~~~~Derived flux\tablenotemark{a}:   & 1.3 Jy & 1.4 Jy \\
Synthesized beam: 
                         & $0\farcs13\times0\farcs10$, p.a. 39
                         & $1\farcs12\times0\farcs93$, p.a. 37 \\
Map rms noise:  & 0.4 mJy & 0.028 mJy \\
\enddata
\tablenotetext{a}{The derived fluxes for 1037-295 were indistinguishable 
for the two epochs (within an estimated uncertainty of 10\%).}
\label{tab:vlaobs}
\end{deluxetable}

\clearpage

\end{document}